\documentclass[twocolumn,tighten]{aastex62}
\usepackage{amssymb, amsmath,framed}
\usepackage{lineno}

\usepackage{bm}
\expandafter\ifx\csname package@font\endcsname\relax\else
 \expandafter\expandafter
 \expandafter\usepackage
 \expandafter\expandafter
 \expandafter{\csname package@font\endcsname}
\fi
\def\bq{\begin{equation}}
\def\eq{\end{equation}}
\def\bqy{\begin{eqnarray}}
\def\eqy{\end{eqnarray}}

\begin{document}
\title{\large{Modeling Martian Atmospheric Losses over Time: Implications for Exoplanetary Climate Evolution and Habitability}}

\correspondingauthor{Chuanfei Dong}
\email{dcfy@princeton.edu}

\author{Chuanfei Dong}
\affiliation{Department of Astrophysical Sciences, Princeton University, Princeton, NJ 08544, USA}
\affiliation{Princeton Center for Heliophysics, Princeton Plasma Physics Laboratory, Princeton University, Princeton, NJ 08540, USA}

\author{Yuni Lee}
\affiliation{NASA Goddard Space Flight Center, Greenbelt, MD 20771, USA}

\author{Yingjuan Ma}
\affiliation{Institute of Geophysics and Planetary Physics, University of California, Los Angeles, CA 90095, USA}

\author{Manasvi Lingam}
\affiliation{Institute for Theory and Computation, Harvard University, Cambridge MA 02138, USA}

\author{Stephen Bougher}
\affiliation{Department of Climate and Space Sciences and Engineering, University of Michigan, Ann Arbor, MI 48109, USA}

\author{Janet Luhmann}
\affiliation{Space Sciences Laboratory, University of California, Berkeley, CA 94720, USA}

\author{Shannon Curry}
\affiliation{Space Sciences Laboratory, University of California, Berkeley, CA 94720, USA}

\author{Gabor Toth}
\affiliation{Department of Climate and Space Sciences and Engineering, University of Michigan, Ann Arbor, MI 48109, USA}

\author{Andrew Nagy}
\affiliation{Department of Climate and Space Sciences and Engineering, University of Michigan, Ann Arbor, MI 48109, USA}

\author{Valeriy Tenishev}
\affiliation{Department of Climate and Space Sciences and Engineering, University of Michigan, Ann Arbor, MI 48109, USA}

\author{Xiaohua Fang}
\affiliation{Laboratory for Atmospheric and Space Physics, University of Colorado, Boulder, CO 80303, USA}

\author{David Mitchell}
\affiliation{Space Sciences Laboratory, University of California, Berkeley, CA 94720, USA}

\author{David Brain}
\affiliation{Laboratory for Atmospheric and Space Physics, University of Colorado, Boulder, CO 80303, USA}

\author{Bruce Jakosky}
\affiliation{Laboratory for Atmospheric and Space Physics, University of Colorado, Boulder, CO 80303, USA}

\begin{abstract}
In this Letter, we make use of sophisticated 3D numerical simulations to assess the extent of atmospheric ion and photochemical losses from Mars over time. We demonstrate that the atmospheric ion escape rates were significantly higher (by more than two orders of magnitude) in the past at $\sim 4$ Ga compared to the present-day value owing to the stronger solar wind and higher ultraviolet fluxes from the young Sun. We found that the photochemical loss of atomic hot oxygen dominates over the total ion loss at the current epoch whilst the atmospheric ion loss is likely much more important at ancient times. We briefly discuss the ensuing implications of high atmospheric ion escape rates in the context of ancient Mars, and exoplanets with similar atmospheric compositions around young solar-type stars and M-dwarfs.
\end{abstract}

\section{Introduction}
Mars has always represented an important target from the standpoint of planetary science \citep{DPL01}, especially on account of its past, and perhaps even its current, biological potential \citep{JP01,Ehl16}. In particular, ancient Mars ($\sim 4$ Ga) has attracted a great deal of attention \citep{Word16} because it may have possessed aqueous environments with water-rock interactions \citep{HG17}, minerals \citep{EE14}, biogenic elements \citep{FD16}, suitable energy sources for prebiotic chemistry \citep{LDF18} and possibly oceans \citep{DH10}; all of these factors could have enhanced the prospects for its habitability. Furthermore, some authors have suggested that the atmospheric composition and conditions of Noachian Mars were fairly similar to Hadean-Archean Earth \citep{McK10} and there is also a non-negligible probability that life could have been transferred from the former to the latter via lithopanspermia.

However, one of the most striking differences between ancient and current Mars is that the former had a thicker atmosphere compared to the present-day value \citep{JP01}, thereby making Noachian Mars potentially more conducive to hosting life. This discrepancy immediately raises the question of how and when the majority of the Martian atmosphere was lost, as well as the channels through which it occurred \citep{brain}. There are compelling observational and theoretical reasons to believe that the majority of atmospheric escape must have occurred early in the planet's geological history \citep{Lammer13,Jak17}, when the extreme ultraviolet (EUV) flux and the solar wind from the Sun were much stronger than today \citep{RGG05,BLK10,DHL17}. Moreover, the Martian dynamo shut down $\sim 4.1$ Ga and Mars currently has only weak crustal magnetic fields \citep{Lil13}. Our understanding of present-day Martian atmospheric escape has improved greatly thanks to observations undertaken by, e.g., the Mars Atmosphere and Volatile EvolutioN mission (MAVEN) \citep{JL15} in conjunction with detailed theoretical modeling \citep{lee15b,Fan17,bougher17,MR17,dong18}.

In this Letter, we will make use of the one-way coupled framework developed by \citet{dong15b} and \citet{lee15a}, known to accurately reproduce MAVEN observations, for studying the ion and photochemical escape rates over the history of Mars while self-consistently accounting for increased EUV and solar wind. The outline of the paper is as follows. In Sec. \ref{SecModel}, we will describe the models and our numerical setup. We follow this up by describing and analyzing our results in Sec. \ref{SecResDisc}. We conclude by summarizing the salient points in Sec. \ref{SecConc}.

\section{Model Descriptions and Setup}\label{SecModel}
Here, we briefly outline the three sophisticated 3D global models for the Martian (i) ionosphere-thermosphere, (ii) exosphere, and (iii) magnetosphere.

We simulate the ionosphere and thermosphere by employing the Mars Global Ionosphere Thermosphere Model (M-GITM) \citep{bougher15}. M-GITM is a 3D ``whole atmosphere'' (ground to exobase) non-hydrostatic model that includes all of the important ion-neutral chemistry and key radiative processes. M-GITM currently solves for neutral and ion densities, as well as neutral temperatures and winds around the globe. In this study, we initialize the Martian atmosphere by using current parameters since it has been shown that both surface pressure \citep{DLMC17} and atmospheric composition \citep{brain} do not have a significant impact on atmospheric escape rates.

Above certain altitudes (i.e., beyond the exobase), the fluid assumption is generally not valid anymore, thus a kinetic model has to be used to model the nearly collisionless exosphere. The dissociative recombination of O$_2^+$ (that not only splits the recombined molecular O$_2$ into atomic O but also gives the resultant atomic O additional kinetic energy) is the most important reaction, primarily responsible for producing the dayside atomic oxygen exosphere. In order to simulate the 3D hot oxygen corona and the associated photochemical escape (i.e., loss of energetic atomic oxygen to space), we use the 3D Mars Adaptive Mesh Particle Simulator (AMPS) that solves the Boltzmann equation in the test-particle mode using the Direct Simulation Monte Carlo (DSMC) method. In Mars AMPS, the ionosphere-thermosphere inputs are taken from M-GITM \citep{lee15a}. Both M-GITM and Mars AMPS operate in the Geographic (GEO) coordinate system. 

Lastly, the 3-D BATS-R-US Mars multifluid MHD (MF-MHD) model starts from 100 km above the Martian surface unlike its Earth counterpart that starts from $\sim 2$-$3$ Earth radii. MF-MHD solves separate continuity, momentum and energy equations for four ion fluids H$^+$, O$^+$, O$_2^+$, CO$_2^+$ \citep{najib11,dong14}. It includes a self-consistent ionosphere and the concomitant photochemistry such as photoionization, charge exchange, electron impact ionization and ion-electron recombination. In order to capture variations in the lower ionosphere, the radial resolution near the inner boundary is set to 5 km. The crustal magnetic fields are implemented based on the 60 degree harmonic expansion model adopted in \citet{dong14}, the strongest of which we set to face nightside in this study for simplicity. For modeling the ionosphere, magnetosphere and the associated atmospheric ion loss, we take the advantage of the existent one-way coupled framework, i.e., the M-GITM thermosphere and the AMPS oxygen exosphere are used as inputs for the MF-MHD model. MF-MHD runs in the Mars-centered Solar Orbital (MSO) coordinate system, where the $+$x-axis points from Mars to the Sun, the $+$z-axis is perpendicular to the Martian orbital plane and points northward, and the y-axis completes the right-hand system. 

We study four cases and the corresponding parameters are listed in Table \ref{table1}. They include the extreme ultraviolet (EUV) strength, the time before present (t$_{BP}$), nominal solar wind density (n$_{sw}$) and velocity  (v$_{sw}$), interplanetary magnetic field (B$_{IMF}$) and the angle associated with an away sector Parker spiral ($\phi_{IMF}$) \citep{RGG05,BLK10}. The reason we halt our analysis at $\sim$ 4 Ga is because little is known of the pre-Noachian period \citep{CH10} and the solar wind parameters are very uncertain during this epoch \citep{Lundin07}.

\begin{table*}
\caption{Input parameters (first six columns) used for the different cases \citep{RGG05,BLK10}, ion escape rates (7th to 9th columns) and photochemical escape rate (last column). Note that 1 EUV (below) refers to the EUV flux received at Mars during the moderate phase of the solar cycle at the current epoch.}
\centering
\begin{tabular}{c|c|c|c|c|c|c|c|c|c}
\hline
\hline
EUV  & t$_{BP}$ (Ga)  & n$_{sw}$ (cm$^{-3}$)  & v$_{sw}$ (km/s) & B$_{IMF}$ (nT) & $\phi_{IMF}$ (degree) & O$^+$ (s$^{-1}$) & O$_2^+$ (s$^{-1}$) & CO$_2^+$ (s$^{-1}$) & O$_{hot}$ (s$^{-1}$)	\\
\hline
1  	&   0.0 	& 2.51  	&  401  	&	3.01		&	58.0 &	1.8$\times$10$^{24}$	&	2.6$\times$10$^{24}$	&	3.6$\times$10$^{23}$	&	2.7$\times$10$^{25}$	\\ 
3  	&   2.77 	& 10.26  	&  578  	&	7.06		&	64.6	&	2.4$\times$10$^{25}$	&	6.6$\times$10$^{24}$	&	1.4$\times$10$^{24}$	&	8.5$\times$10$^{25}$	\\ 
6  	&   3.57 	& 24.75  	&  726 	&	12.17	&	68.2	&	2.4$\times$10$^{26}$	&	9.3$\times$10$^{24}$	&	2.7$\times$10$^{24}$	&	9.9$\times$10$^{25}$	\\  
10  	&   3.93 	& 46.99  	&  858	&	18.16	&	70.5	&	1.1$\times$10$^{27}$	&	1.2$\times$10$^{25}$	&	4.1$\times$10$^{24}$	&	1.0$\times$10$^{26}$	\\ 
\hline
\end{tabular}\label{table1}
\end{table*}

\section{Results and Discussion}\label{SecResDisc}
\begin{figure*}
\centering
\noindent
\includegraphics[width=42pc,angle=0]{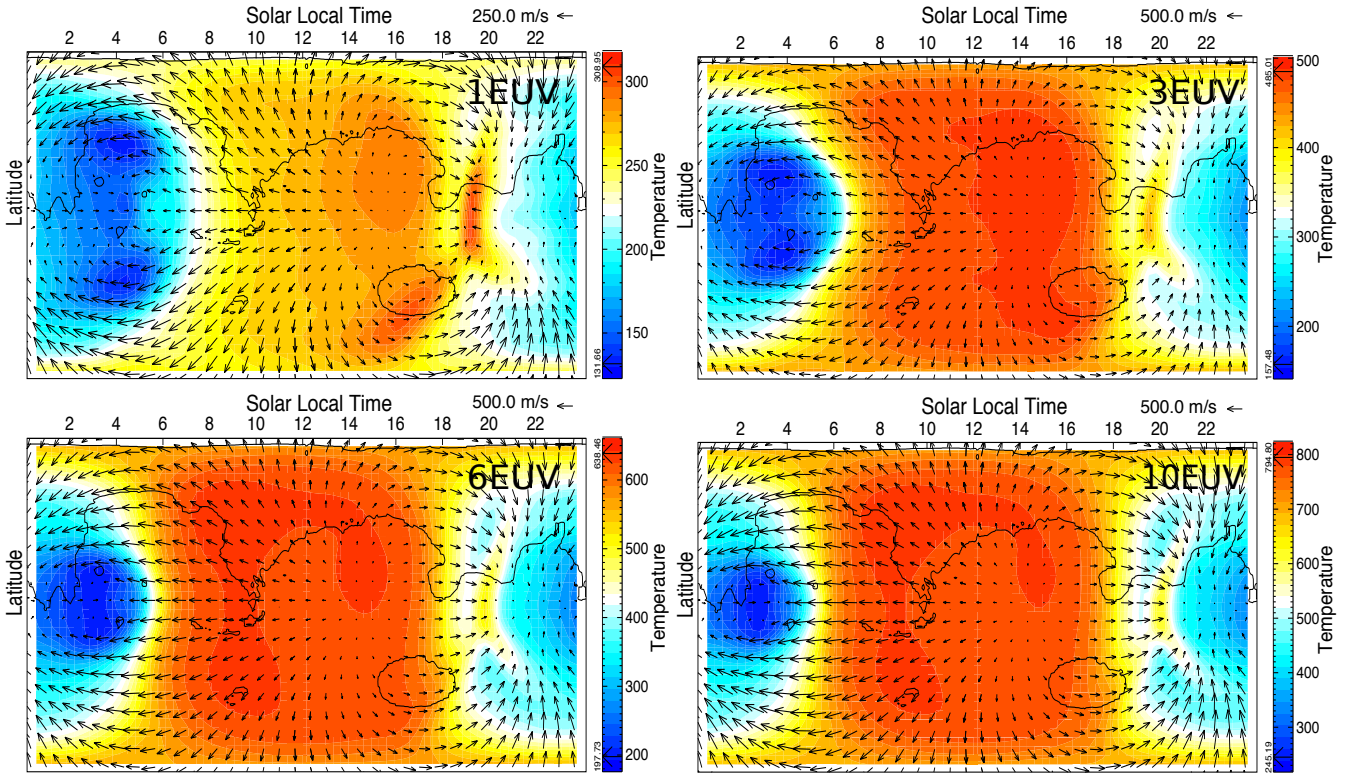}
\caption{Color contours of temperature (in K) at $\sim$ 200 km for 1, 3, 6 and 10 EUV under equinox conditions (that approximately represents an average over one Mars orbit). The arrows in each panel indicate the relative magnitude (reference is given in top right corner) and the direction of the horizontal winds. The vertical axes (i.e. latitude) range between $-$90$^{\circ}$ and 90$^{\circ}$. Note that the colorbar varies in different panels.}
\label{mgitm}
\end{figure*}

Figure \ref{mgitm} depicts the temperature and winds of the Martian thermosphere at $\sim$ 200 km for equinox conditions. An inspection of Figure \ref{mgitm} reveals that a high EUV flux is correlated with a hotter thermosphere. Therefore, the EUV heating of the thermosphere is self-consistently computed, which is very important for deriving the atmospheric ion and photochemical losses.   

\begin{figure}
\centering
\noindent
\includegraphics[width=20pc,angle=0]{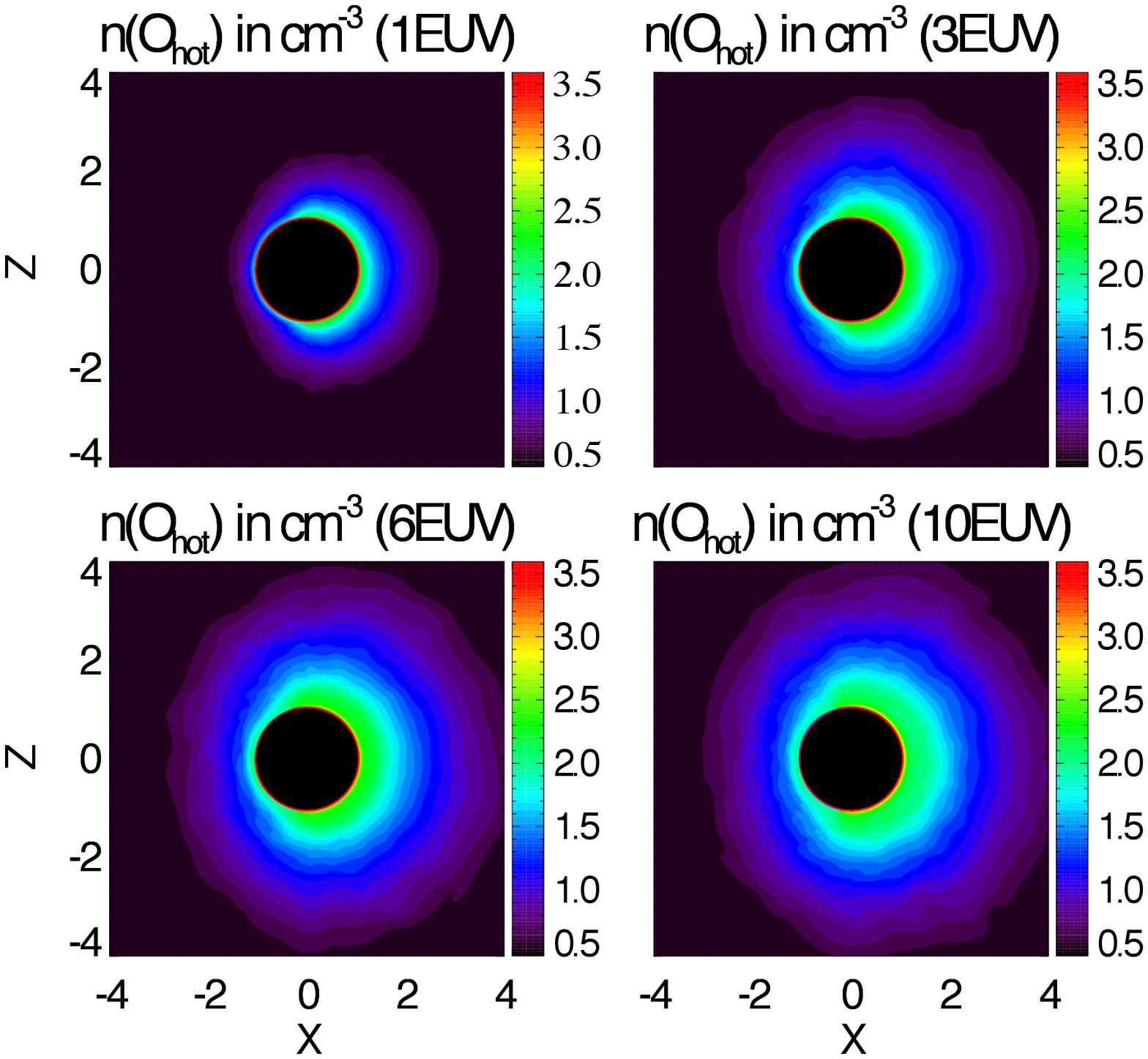}
\caption{Comparison of the hot oxygen density (in cm$^{-3}$) distribution in the $x$-$z$ meridian plane for different EUV cases using a logarithmic scale.}
\label{amps}
\end{figure}

Figure \ref{amps} shows the atomic hot oxygen density distribution in the meridian plane from AMPS based on the M-GITM input. The presented asymmetry in the hot oxygen density distribution is a result of higher O$_2^+$ abundance on the dayside than nightside. Compared to the current epoch with relatively low EUV flux, ancient Mars had a more intensive and extensive oxygen corona resulting from the enhanced O$_2^+$ density at higher EUV flux.

\begin{figure}
\centering
\noindent
\includegraphics[width=20pc,angle=0]{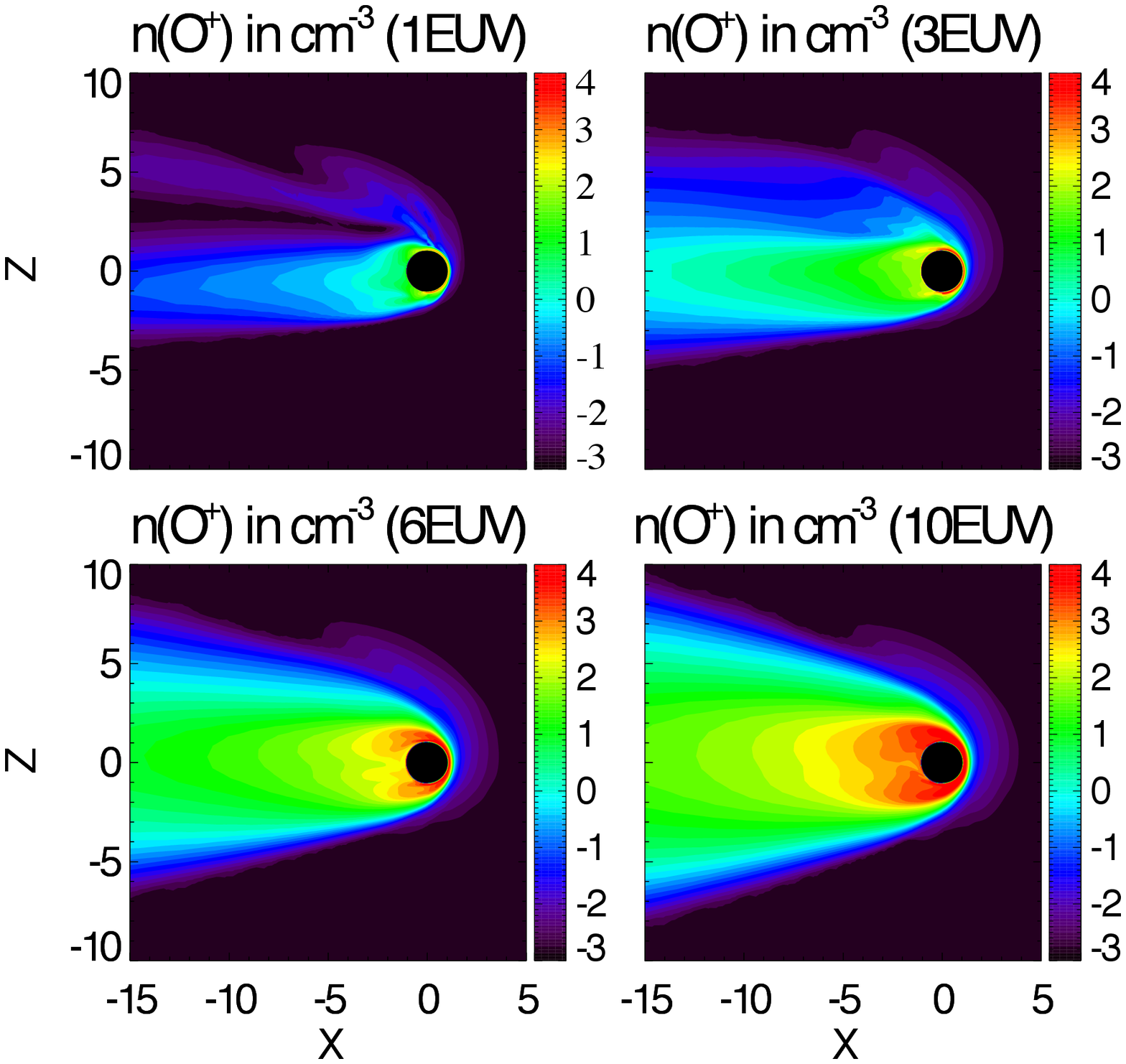}
\caption{Logarithmic scale contour plots of the O$^+$ ion density (in cm$^{-3}$) in the $x$-$z$ meridian plane for different EUV and solar wind cases.}
\label{mfmhd}
\end{figure}

In Figure \ref{mfmhd}, we present the MF-MHD calculation of O$^+$ ion escaping from the planet. One of the features of the MF-MHD model is that it captures the asymmetric ion escape plume, resulting from the Lorentz force term in the individual ion momentum equations \citep{najib11,dong14}, as observed by MAVEN \citep{DF17}. The asymmetric ion escape plume becomes less evident at earlier epochs because of both the extended corona and the smaller ion gyroradius ($\propto$ v$_{sw}$/B$_{y}$) in this period. As seen from Figure \ref{mfmhd}, more O$^+$ ions escape from the planet at higher EUV and stronger solar wind.

\begin{figure}
\centering
\noindent
\includegraphics[width=20pc,angle=0]{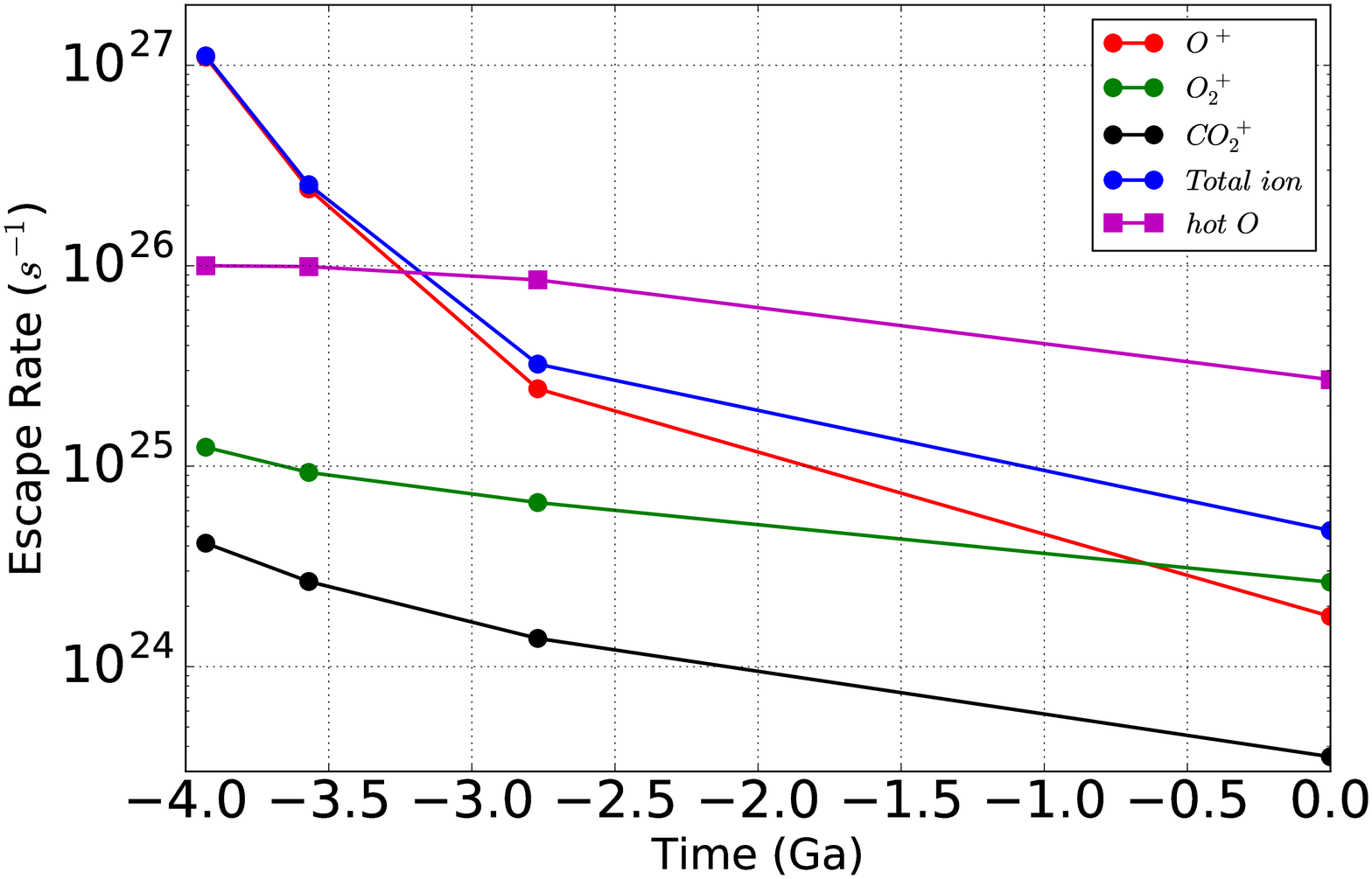}
\caption{Calculated ion and photochemical escape rates over the Martian history.}
\label{ionloss}
\end{figure}

We list the atmospheric ion and photochemical escape rates in Table \ref{table1} and plot them in Figure \ref{ionloss}; see also Fig. 4 of \citet{LJZ92} and \citet{CLL07} where similar calculations were undertaken based on less comprehensive methods. The calculated atmospheric escape rates in Table \ref{table1} are consistent with the density contour plots illustrated in Figures \ref{amps} and \ref{mfmhd}. It is noteworthy that the difference in photochemical loss between the 6 EUV and 10 EUV cases is nearly indistinguishable. The underlying reason is that the enhanced collision probability between hot oxygen and thermal species in the extended thermosphere can deflect hot/energetic particles more efficiently and thus decreases the escape probability of hot O \citep{zhao2015}. Interestingly, the photochemical escape rate of atomic hot oxygen dominates over ion losses at the current epoch whilst the atmospheric ion escape rate becomes an order of magnitude larger than photochemical loss at ancient times, indicating that atmospheric losses are primarily controlled by ion escape for early Mars. In addition, compared to molecular ion species (O$_2^+$ and CO$_2^+$), O$^+$ is the dominant ion depleted at early epochs.

An analytic estimate of the total atmospheric escape rate $\dot{N}$ from weakly magnetized planets due to the erosion by the solar wind is via $\dot{N} \propto \left({R_p}/{a}\right)^2 \dot{M}_\star$, where $R_p$ and $a$ are the planet's radius and semi-major axis respectively, while $\dot{M}_\star$ denotes the solar mass-loss rate \citep{ZSR10,DJL18}; also see, e.g., \citet{Cravens17b} for related analyses. We end up with $\dot{N} \propto t^{-2.33}$, where $t$ is the age of the star, because $\dot{M}_\star$ exhibits this time dependence for solar-type stars \citep{WMZ05}. The photochemical atmospheric loss will be primarily driven by the EUV flux. Hence, if the photochemical escape rate $\dot{N}_O$ is proportional to the EUV flux $\Phi_{EUV}$ \citep{Cravens17a}, we obtain $\dot{N}_O \propto t^{-1.19}$ because $\Phi_{EUV}$ displays this time dependence \citep{RGG05}. We find that the analytical trends are in good agreement with the numerical simulations at later epochs ($t > 1$ Gyr) but are less accurate for ancient Mars.

From Figure \ref{ionloss}, it can be seen that the total ion escape rate was $\gtrsim 100$ times higher than today at $\sim 4$ Ga. Hence, our results are consistent with Mars having lost much of its atmosphere early in its history, leading to the Martian climate changing from a warm and wet environment in the past to the desiccated, frigid and thus inhospitable one documented today. Our simulations indicate that the total photochemical and ion atmospheric losses over the span $\sim 0$-$4$ Ga are approximately equal to each other, and their sum amounts to $\sim 0.1$ bar being lost over this duration. If we assume that the oxygen lost through a combination of ion and photochemical escape mechanisms was originally derived from surface water, we find $\sim 3.8 \times 10^{17}$ kg of water has been lost from Mars between $0$ to $\sim 4$ Ga; this mass corresponds to a global surface depth of $\sim 2.6$ m (the depth will be greater if the water bodies were more localized). 
The calculations do not include other potentially important loss processes such as sputtering; therefore, it provides a lower limit on the escape rates. Our result is more conservative compared to earlier studies \citep{ZLB93,Luh97,CLL07,VBT10} that predicted $\mathcal{O}(10)$ m of water was depleted over Martian history.

Before proceeding further, recall that our analytic estimates were expressible as $\dot{N} \propto t^{-\alpha}$ and $\dot{N}_O \propto t^{-\beta}$ with $\alpha \approx 2.33$ and $\beta \approx 1.19$. Our choice of these values was motivated by the fact that the solar wind parameters used in the numerical simulations \citep{BLK10} were consistent with \citet{WMZ05}, and the ancient EUV fluxes were based on \citet{RGG05}. In actuality, the exponent $\alpha$ is not tightly constrained and recent studies favor $\alpha \lesssim 1$ \citep{CS11,JGB15}. Similarly, other methods for calculating the EUV flux over time - based, for instance, on X-ray \citep{CGU15} and Ly-$\alpha$ \citep{LFF14} emission - lead to different values of $\beta$. Since the integrated mass lost over the duration $0$ to $\sim 4$ Ga (due to ion loss mechanisms) is proportional to $\left(\alpha-1\right)^{-1}\left[(23/3)^{\alpha-1}-1\right]$, we find that the overall estimated value changes at most by one order of magnitude for $\alpha \in \left(0.5,2.5\right)$; the factor of $23/3$ occurs because the current age of $4.6$ Gyr ($0$ Ga) is divided by $0.6$ Gyr ($\sim 4$ Ga). The same conclusion also holds true for photochemical escape since the corresponding mass lost is found via the substitution $\alpha \rightarrow \beta$. Hence, it seems reasonable to conclude that the basic conclusions in this Section are not greatly altered if the values of $\alpha$ or $\beta$ are changed.

\section{Conclusion}\label{SecConc}
In our solar system, Mars represents a classic example of a planet where planetary habitability has been unambiguously affected by atmospheric losses. In this Letter, we have studied the atmospheric ion and photochemical escape rates from Mars over time. We found that the atmospheric ion escape rates vary significantly over the planet's history, ranging from $\mathcal{O}\left(10^{27}\right)\,\mathrm{s}^{-1}$ at $\sim 4$ Ga to $\mathcal{O}\left(10^{24}\right)\,\mathrm{s}^{-1}$ in the present epoch. The corresponding photochemical escape rate lies between $\mathcal{O}\left(10^{26}\right)\,\mathrm{s}^{-1}$ at $\sim 4$ Ga and $\mathcal{O}\left(10^{25}\right)\,\mathrm{s}^{-1}$ today. Therefore, our simulations are consistent with the idea that Mars could have transitioned from having a thick atmosphere and global water bodies to its current state with a tenuous atmosphere and arid conditions quite early in its history. The total atmospheric loss over time predicted by simulations may, perhaps, be tested against observations (to some degree) by using isotope ratios, since the lighter ions are picked up preferentially compared to the heavier ions, akin to the method used by \citet{Jak17}. However, we caution the reader that the uncertainties involved with the solar wind and EUV flux increase as we move towards more ancient epochs, implying that our conclusions concerning atmospheric losses over time will also be subject to a certain degree of variability. 

Our results also have implications for the rapidly expanding domain of exoplanets if one views Mars as a prototype for small rocky exoplanets. Water can be lost from the atmospheres of exoplanets in the habitable zone (HZ) of M-dwarfs over relatively fast timescales \citep{BSO17}, compared to heavier molecules (e.g. CO$_2$). Since the total number of rocky exoplanets in the HZ of M-dwarfs is expected to be $\sim 10^{10}$ \citep{DC15}, the possibility of exoplanets with atmospheric compositions similar to Venus and Mars cannot be ruled out. Our work demonstrates that such exoplanets, as well as those around young solar-type stars, could be subjected to high atmospheric escape rates early in their history. For exoplanets orbiting M-dwarfs, the situation could be even worse due to the more intense particle and radiation environments that exoplanets experience in their close-in HZs. If the escape rates scale as $1/a^{2}$, it is possible for $\sim 100$ bars to be lost from a Mars-like exoplanet in the HZ of an M-dwarf of mass $\sim 0.1\,M_\odot$ over a span of $\sim 4.0$ Gyr. Equivalently, this corresponds to a global water depth of $\sim 2.6$ km being depleted if the source of atmospheric oxygen was surface water. In turn, if their atmospheres and oceans end up being altogether depleted over sub-Gyr timescales, this could lead to detrimental effects insofar their habitability is concerned \citep{DLMC17,DJL18,LL18}. 

Thus, from a broader perspective, our work demonstrates that atmospheric loss is not static but dynamical in nature and that high escape rates will typically occur early in the host star's history. It is therefore necessary to take this time-dependence into account when modeling atmospheric loss from early Mars, and Mars-like exoplanets in the future. We also expect to incorporate the impact of extreme space weather events that are highly frequent on young and/or low-mass stars in our future study \citep{DHL17} to model the atmospheric loss and evolution of early Mars/Mars-like planets.

\acknowledgments
This research was supported by NASA grant NNH10CC04C through MAVEN Project and NASA grant 80NSSC18K0288. Resources supporting this work were provided by the NASA High-End Computing (HEC) Program through the NASA Advanced Supercomputing (NAS) Division at Ames Research Center. The Space Weather Modeling Framework that contains the BATS-R-US code used in this study is publicly available from http://csem.engin.umich.edu/tools/swmf. For distribution of the model results used in this study, please contact the corresponding author. 


\end{document}